\documentstyle[aps,twocolumn]{revtex}

\begin{document}
\author{Li-Xiang Cen$^1$\thanks{%
Electronic address: cenlx@red.semi.ac.cn}, Nan-Jian Wu$^1$, Fu-Hua Yang$^1$,
and Jun-Hong An$^2$}
\address{$^1$NLSM, Institute of Semiconductors, Chinese Academy of Sciences,
Beijing 100083, P.R. China,\\
$^2$Department of Modern Physics, Lanzhou University, Lanzhou 730000,
P.R.China}
\title{Local transformation of mixed states of two qubits to Bell diagonal states}
\maketitle

\begin{abstract}
The optimal entanglement manipulation for a single copy of mixed states of
two qubits is to transform it to a Bell diagonal state. In this paper we
derive an explicit form of the local operation that can realize such a
transformation. The result obtained is universal for arbitrary entangled
two-qubit states and it discloses that the corresponding local filter is not
unique for density matrices with rank $n=2$ and can be exclusively
determined for that with $n=3$ and $4$. As illustrations, a four-parameters
family of mixed states are explored, the local filter as well as the
transformation probability are given explicitly, which verify the validity
of the general result.
\end{abstract}


Entanglement manipulation is an important issue in the studies of quantum
information theory. On the one hand, it is related to the basic problem of
which tasks one can accomplish with a given resource of entanglement \cite
{bennett1}-\cite{monotone}. On the other hand, since most applications of
quantum information theory require the maximally entangled state for
faithful transmission of quantum data, it is necessary to develop the
special technique of entanglement manipulation which uses local quantum
operations and classical communication (LOCC) to produce states with the
amount of entanglement as great as possible from partly entangled states 
\cite{procrus,bennett2}.

There has been much attention recently concerning the entanglement
manipulation of a single copy of mixed two-qubit states \cite{gisin}-\cite
{thew}. A basic measure of entanglement of mixed states is the entanglement
of formation, which is intended to quantify the entanglement resources
needed to create the states \cite{bennettm}. Given a two-qubit density
matrix $\rho $, the entanglement of formation is defined as 
\begin{equation}
E(\rho )=\min \sum_ip_iE(\psi _i),  \label{entdef}
\end{equation}
where the minimum is taken over all decompositions of $\rho $ into pure
states 
\begin{equation}
\rho =\sum_ip_i|\psi _i\rangle \langle \psi _i|  \label{decomp}
\end{equation}
and $E(\psi _i)$ is the entropy of entanglement of the pure state $|\psi
_i\rangle $. Although the expected entanglement cannot be increased 
by any LOCC actions, one can gamble, with risk of failure for some 
probabilities, on obtaining states with more entanglement from less 
entangled states \cite{procrus}. It was proven by Kent etc. \cite
{kent} that under LOCC operations, the entanglement of formation of a single
copy of two-qubit states exists an upper bound, the maximum extractable
entanglement, and the corresponding state is a Bell diagonal state which is
unique up to local unitary transformations. In this stage, a critical
problem is to calculate the detailed LOCC operations and determine the
probability of success of such a transformation for given two-qubit states.

The local operation on the two-qubit state can be studied in a real matrix
parameterization representation of the state, i.e., the Horodecki
representation \cite{horod}. In a recent paper \cite{verst}, it was shown
that local operators performed on two qubits correspond to left and right
multiplication by a Lorentz matrix, respectively, and useful results have
been derived from such a point of view. However, until now, the above
problem is still open and the distinct form of the LOCC actions that realize
the optimal transformation is lack. In the presented paper we investigate
this problem in a particular representation, the ''Wootters representation''
of two-qubit states \cite{wootters1}. We shall show, by introducing an
associated operator---a squared quantity of the local filter, the explicit
form of the optimal transformation as well as the probability can be derived
for arbitrary entangled two-qubit states. Moreover, the result displays that
the corresponding local filter is not unique for density matrices with rank $n=2$
and can be exclusively determined for that with $n=3$ and $4$. As illustrations,
a four-parameters family of mixed states shall be explored, the local filter
and the transformation probability are given explicitly, so that the validity of
the general result is verified.

{\it Main result}.---Consider the local operations on a single copy of
two-qubit states. Recall that the local operator can be written as a unitary
transformation multiplying a Hermitian filtering operator \cite{purify}. One
can, without loss of generality, describe the transformation as 
\begin{equation}
\rho \rightarrow \rho ^{\prime }=\frac{f_A\otimes f_B\rho f_A\otimes f_B}{%
tr(f_A^2\otimes f_B^2\rho )}.  \label{gener}
\end{equation}
Here, we have neglected the local unitary transformation since they shall
preserve the entanglement. The local filters $f_A$ and $f_B$ are given by 
\cite{purify,cen} 
\begin{eqnarray}
f_A=\frac 1{1+a}(I_2+a{\bf m}\cdot {\bf \sigma }),  \label{filter1} \\
f_B=\frac 1{1+b}(I_2+b{\bf n}\cdot {\bf \sigma }),  \label{filter}
\end{eqnarray}
where ${\bf m}$, ${\bf n}$ are unit vectors and the parameters $a$ and $b$
satisfy $0\leq a\leq 1$, $0\leq b\leq 1$. Above expressions have excluded
the trivial cases of the filters with the determinants $\det
(I_2-f_{A(B)}^2)>0$ which differentiate from the operators (\ref{filter1})
and (\ref{filter}) only by a coefficient less than $1$. Introduce an
associated operator of the local filter $f_A\otimes f_B$%
\begin{equation}
F_{AB}\equiv \frac{f_A^2\otimes f_B^2}{\det f_A\det f_B}.  \label{form}
\end{equation}
This quadratic form $F_{AB}$ is equivalent to the local filter $f_A\otimes
f_B $ in a sense that once $F_{AB}$ is provided, the filtration parameters $%
a $, $b$, ${\bf m}$, and ${\bf n}$ can be uniquely determined from Eqs. (\ref
{filter1})-(\ref{form}). It can be easily checked, from the relation $%
f_{A(B)}\tilde{f}_{A(B)}=\det f_{A(B)}I_2$, that the operator $F_{AB}$ has
the property 
\begin{equation}
F_{AB}\tilde{F}_{AB}=I_2\otimes I_2,  \label{norm}
\end{equation}
where the tilde ''\symbol{126}'' of an operator is defined as $\tilde{F}%
_{AB}\equiv \sigma _2\otimes \sigma _2F_{AB}^{*}\sigma _2\otimes \sigma _2$.
In the following, we shall determine the detailed form of $F_{AB}$ for the
optimal entanglement manipulation which transforms the two-qubit state $\rho 
$ into a Bell diagonal state.

We make use of the so-called ''Wootters' representation'', i.e., the
particular decomposition of the two-qubit state \cite{wootters1} 
\begin{equation}
\rho =\sum_{i=1}^n|x_i\rangle \langle x_i|.  \label{represen}
\end{equation}
Here, $n\leq 4$ is the rank of $\rho $. The states $|x_i\rangle $ satisfy 
\begin{equation}
\langle x_i|\tilde{x}_j\rangle =\lambda _i\delta _{ij},\ \ |\tilde{x}%
_i\rangle =\sigma _2\otimes \sigma _2|x_i^{*}\rangle ,  \label{inner}
\end{equation}
and the $\lambda _i$s are the eigenvalues, in decreasing order, of the
Hermitian matrix $R(\rho )\equiv \sqrt{\sqrt{\rho }\tilde{\rho}\sqrt{\rho }}$%
. A combination of these symbols defines the concurrence for the entangled
state $\rho $: $C(\rho )=\lambda _1-\lambda _2-\lambda _3-\lambda _4$. It
turns out that the above representation is very useful to determine the
operator $F_{AB}$. Namely, we have

Proposition.---An entangled two-qubit state $\rho $ given by (\ref{represen}%
) with $\lambda _n>0$ can be transformed to a Bell diagonal state by the
local filter $f_A\otimes f_B$ if and only if the associated operator $F_{AB}$
satisfies 
\begin{equation}
\langle x_i|F_{AB}|x_i\rangle =\langle x_i|\tilde{x}_i\rangle ,\ \
i=1,\cdots ,n.  \label{condit}
\end{equation}

Relation (\ref{condit}) implies that the normalized states 
\begin{equation}
|E_i\rangle \equiv \frac{f_A\otimes f_B|x_i\rangle }{\sqrt{\langle
x_i|f_A^2\otimes f_B^2|x_i\rangle }},\ \ i=1,\cdots ,n  \label{define}
\end{equation}
have the ''tilde inner products'' $\langle E_i|\tilde{E}_i\rangle =1$, thus $%
|E_i\rangle =|\tilde{E}_i\rangle $. This yields explicitly 
\begin{equation}
\langle x_i|F_{AB}|x_j\rangle =\langle x_i|\tilde{x}_j\rangle =0,\ \ i\neq j.
\label{condit1}
\end{equation}

The sufficiency of the proposition can be shown directly. Following the fact
that the states $|E_i\rangle $ ($i=1,\cdots ,n$) have the concurrence $1$,
that is, they are a set of maximally entangled states of qubits A and B, one
can conclude that the transformed state $\rho ^{\prime }$ has completely
random local density matrices $tr_A\rho ^{\prime }=tr_B\rho ^{\prime }=\frac %
12I_2$, thus is Bell diagonal. To prove the converse statement, we use the
fact that the Bell diagonal state has the property \cite{cen} 
\begin{equation}
trR(\rho )=\sum_{i=1}^n\lambda _i=1.  \label{trace}
\end{equation}
Now suppose the transformed state 
\begin{equation}
\rho ^{\prime }=\frac{\sum_{i=1}^nf_A\otimes f_B|x_i\rangle \langle
x_i|f_A\otimes f_B}{tr(\rho f_A^2\otimes f_B^2)}=\sum_{i=1}^np_i|E_i\rangle
\langle E_i|  \label{transtate}
\end{equation}
is Bell diagonal. Here, $|E_i\rangle $ is given by Eq. (\ref{define}) and $%
p_i$ is the probability of it in the decomposition. Direct observation can
be found that the states $|E_i\rangle $ ($i=1,\cdots ,n$) have tilde inner
products $\langle E_i|\tilde{E}_j\rangle =\langle E_i|\tilde{E} _i\rangle
\delta _{ij}$ with $\langle E_i|\tilde{E}_i\rangle $ positive real numbers.
Thus we have 
\begin{equation}
\sum_{i=1}^ntrR(\rho ^{\prime })=\sum_{i=1}^np_i\langle E_i|\tilde{E}
_i\rangle =1.  \label{trrel}
\end{equation}
This yields that $\langle E_i|\tilde{E}_i\rangle =1$ for all $i=1,\cdots ,n$%
, then the relation (\ref{condit}).

The analysis presented above also indicates that the two-qubit states with $%
\lambda _n=0$ cannot be transformed into Bell diagonal states by any local
actions efficiently. Note that $\lambda _n=0$ can only be possible when $n=3$
and $n=2$ \cite{note}. In such cases, $|x_n\rangle $ is a factorizable state
of qubits A and B, and cannot be transformed to the maximally entangled
state (corresponding to $|E_n\rangle $ in the above description) by any
local filtering operations. The best one can do, is to reduce the
probability of this term to the infinitesimal. This process, as has been
pointed out in Ref. \cite{verst}, transforms the state asymptotically to a
Bell diagonal state with a lower rank and an infinitesimal probability.

Equations (\ref{condit}) and (\ref{condit1}) suggest a distinct way to
construct the associated local operator $F_{AB}$. Knowing that the rank $n=1$
corresponds to pure states and they can be purified by the Procrustean
protocol \cite{procrus}, we discuss in the following the cases of $n=4$, $3$%
, and $2$, respectively.

1. $n=4$. In this case, the states \{$|x_i\rangle ,i=1,\cdots ,4$\} compose
a complete set of non-orthogonal bases of the $4$-dimensional Hilbert space
of the two qubits. Equations (\ref{condit}) and (\ref{condit1}) thus provide
a clear representation of the operator $F_{AB}$ in these bases. Recalling \{$%
|\tilde{x}_i\rangle ,i=1,\cdots ,4$\} the set of biorthogonal bases to \{$%
|x_i\rangle ,i=1,\cdots ,4$\}, we can present explicitly 
\begin{equation}
F_{AB}=\sum_{i=1}^4\frac{|\tilde{x}_i\rangle \langle \tilde{x}_i|}{\langle
x_i|\tilde{x}_i\rangle }.  \label{operator}
\end{equation}
To derive detailed expressions for the local filters $f_A$ and $f_B$, we
need to factorize the operator given above. This shall be shown later in
this section.

2. $n=3$. Apparently, there exists a unique state $|x_4\rangle $ in the $4$%
-dimensional Hilbert space which is orthogonal to the three
linearly-independent states \{$|\tilde{x}_i\rangle ,i=1,2,3$\}. Noticing the
relation $|E_i\rangle =|\tilde{E}_i\rangle $ ($i=1,2,3$), one can find that 
\begin{equation}
\langle x_4|F_{AB}|x_i\rangle =\langle x_i|F_{AB}|x_4\rangle =0,\ \ \
i=1,2,3.  \label{gauge}
\end{equation}
Suppose $\langle x_4|F_{AB}|x_4\rangle =t_{44}$. The operator $F_{AB}$ can
be expressed as 
\begin{equation}
F_{AB}=\sum_{i=1}^3\frac{|\tilde{x}_i\rangle \langle \tilde{x}_i|}{\langle
x_i|\tilde{x}_i\rangle }+t_{44}\frac{|\tilde{x}_4\rangle \langle \tilde{x}_4|%
}{\langle x_4|\tilde{x}_4\rangle ^2}.  \label{form1}
\end{equation}
Substituting this expression into Eq. (\ref{norm}), one obtains $t_{44}=$ $%
\langle x_4|\tilde{x}_4\rangle $. Thus in this case the local operator $%
F_{AB}$ is also unique and can be expressed uniformly as (\ref{operator}).

3. $n=2$. Similarly, provided the states \{$|x_1\rangle $, $|x_2\rangle $\},
one can easily construct the complete set \{$|x_i\rangle ,i=1,\cdots ,4$\}
in the $4$-dimensional Hilbert space of the two qubits. In this case, the
states $|x_3\rangle $ and $|x_4\rangle $ can be chosen variously. Suppose \{$%
|x_i\rangle ,i=1,\cdots ,4$\} is one of the selected sets satisfying (\ref
{inner}). Based on an analysis similar to the previous case, we can present
the form of the Hermitian operator $F_{AB}$ as 
\begin{eqnarray}
F_{AB}&=&\sum_{i=1}^2\frac{|\tilde{x}_i\rangle \langle \tilde{x}_i|}{\langle
x_i|\tilde{x}_i\rangle }+\sum_{i=3}^4t_{ii}\frac{|\tilde{x}_i\rangle \langle 
\tilde{x}_i|}{\langle x_i|\tilde{x}_i\rangle ^2}
\nonumber\\
&+&t_{34}(\frac{|\tilde{x}
_3\rangle \langle \tilde{x}_4|}{\langle x_3|\tilde{x}_3\rangle \langle x_4|
\tilde{x}_4\rangle }+\frac{|\tilde{x}_4\rangle \langle \tilde{x}_3|}{\langle
x_3|\tilde{x}_3\rangle \langle x_4|\tilde{x}_4\rangle }).  \label{case3}
\end{eqnarray}
Direct calculations from the restriction (\ref{norm}) can show that 
\begin{equation}
t_{34}=0,\ \ t_{33}=\langle x_3|\tilde{x}_3\rangle ,\ \ t_{44}=\langle x_4|%
\tilde{x}_4\rangle .  \label{tcase3}
\end{equation}
This yields again the form of (\ref{operator}). The operator $F_{AB}$ now is
not unique due to different selections of the states \{$|x_3\rangle
,|x_4\rangle $\}. Specifically, they can be obtained through constructing
the states as 
\begin{eqnarray}
|x_3^{\prime }\rangle =\frac{c_1|x_3\rangle }{\sqrt{\langle x_3|\tilde{x}%
_3\rangle }}+\frac{d_1|x_4\rangle }{\sqrt{\langle x_4|\tilde{x}_4\rangle }},
\nonumber \\
|x_4^{\prime }\rangle =\frac{c_2|x_3\rangle }{\sqrt{\langle x_3|\tilde{x}%
_3\rangle }}+\frac{d_2|x_4\rangle }{\sqrt{\langle x_4|\tilde{x}_4\rangle }},
\label{recons}
\end{eqnarray}
with the complex parameters $c_i$ and $d_i$ satisfying 
\begin{eqnarray}
c_1^{*}c_2^{*}+d_1^{*}d_2^{*}=0,  \nonumber \\
(c_1^{*})^2+(d_1^{*})^2=1,  \nonumber \\
(c_2^{*})^2+(d_2^{*})^2=1.  \label{multir}
\end{eqnarray}

We now need to show the operator $F_{AB}$ presented in Eq. (\ref{operator})
can always be factorized as a product form of local operators of qubits A
and B. To this end, we subnormalize the states \{$|\tilde{x}_i\rangle,
i=1,\cdots,4$\} such that 
\begin{equation}
F_{AB}=\sum_{i=1}^4|y_i\rangle \langle y_i|,\ \ |y_i\rangle =\frac{|\tilde{x}%
_i\rangle }{\sqrt{\langle x_i|\tilde{x}_i\rangle }}.  \label{yexpre}
\end{equation}
The states \{$|y_i\rangle $\} have orthogonal normalized tilde inner
products 
\begin{equation}
\langle y_i|\tilde{y}_j\rangle =\delta _{ij},\ \ i,j=1,\cdots ,4.
\label{normt}
\end{equation}
Construct a new decomposition of the operator $F_{AB}=\sum_{i=1}^4|z_i
\rangle \langle z_i|$ with 
\begin{eqnarray}
|z_1\rangle =\frac 12[(|y_1\rangle +|y_2\rangle )+i(|y_3\rangle +|y_4\rangle
)],  \nonumber \\
|z_2\rangle =\frac 12[(|y_1\rangle +|y_2\rangle )-i(|y_3\rangle +|y_4\rangle
)],  \nonumber \\
|z_3\rangle =\frac 12[(|y_1\rangle -|y_2\rangle )+i(|y_3\rangle -|y_4\rangle
)],  \nonumber \\
|z_4\rangle =\frac 12[(|y_1\rangle -|y_2\rangle )-i(|y_3\rangle -|y_4\rangle
)].  \label{newset}
\end{eqnarray}
One can check directly that the four states $|z_i\rangle $ ($i=1,\cdots ,4$)
have the concurrence zero, thus are factorizable states of qubits A and B.
In detail, there are 
\begin{eqnarray}
\langle z_1|\tilde{z}_j\rangle =\delta _{2j},\ \ \ \langle z_2|\tilde{z}%
_j\rangle =\delta _{1j},  \nonumber \\
\langle z_3|\tilde{z}_j\rangle =\delta _{4j},\ \ \ \langle z_4|\tilde{z}%
_j\rangle =\delta _{3j.}  \label{zconc}
\end{eqnarray}
Now we show that these relations together with the linearly-independent
property of the states \{$|z_i\rangle ,i=1,\cdots ,4$\} can predict that the
operator $F_{AB}$ takes a factorizable form.

Let us first assume that 
\begin{equation}
|z_1\rangle =\alpha _1|a_1b_1\rangle ,\ \ |z_2\rangle =\alpha
_2|a_2b_2\rangle ,  \label{zexpre}
\end{equation}
where the local basis $|a_i\rangle $ and $|b_i\rangle $ are normalized.
According to relations $\langle z_1|\tilde{z}_1\rangle =\langle z_3|\tilde{z}%
_3\rangle =\langle z_1|\tilde{z}_3\rangle =0$, one can check that an
arbitrary superposition of the two factorizable states $|z_1\rangle$ and $%
|z_3\rangle$ shall still be a factorizable state of qubits A and B. So it
can be concluded that the state $|z_3\rangle $ must take a form of $%
|z_3\rangle =\alpha _3|a_1b_3\rangle $ or $|z_3\rangle =\alpha
_3|a_3b_1\rangle $. Note there are the same relations for the states \{$%
|z_2\rangle ,|z_3\rangle $\}. Combining these two restrictions gives that
the state $|z_3\rangle $ must take the form 
\begin{equation}
|z_3\rangle =\alpha _3|a_1b_2\rangle  \label{z31}
\end{equation}
or 
\begin{equation}
|z_3\rangle =\alpha _3|a_2b_1\rangle .  \label{z32}
\end{equation}
Correspondingly, the similar analysis is applicable to the states \{$%
|z_1\rangle ,|z_2\rangle ,|z_4\rangle $\}. Knowing that the four states $%
|z_i\rangle $ ($i=1,\cdots ,4$) are linearly independent, we can obtain the
form of the state $|z_4\rangle $: 
\begin{equation}
|z_4\rangle =\alpha _4|a_2b_1\rangle  \label{equa1}
\end{equation}
or 
\begin{equation}
|z_4\rangle =\alpha _4|a_1b_2\rangle  \label{equa2}
\end{equation}
corresponding to the two cases of (\ref{z31}) and (\ref{z32}), respectively.
Finally, according to the relation $\langle z_1|\tilde{z}_2\rangle =\langle
z_3|\tilde{z}_4\rangle =1$, the complex coefficients of the states must
satisfy $|\alpha _1\alpha _2|=|\alpha _3\alpha _4|$ for both of these two
cases. Now it can be shown directly that the operator $F_{AB}$ takes a
factorizable form of local filters of qubits A and B.

Up to now, we have presented a distinct expression for the associated
operator $F_{AB}$ and shown that the local filters $f_A$ and $f_B$ can be
obtained through factorizing $F_{AB}$ straightforwardly. To complete the
argument, we need to calculate the probability $P_f$ that the LOCC action
succeeds. Directly, 
\begin{equation}
P_f=tr(\rho f_A^2\otimes f_B^2)=\det f_A\det f_B\sum_{i=1}^n\langle
x_i|F_{AB}|x_i\rangle .  \label{proba}
\end{equation}
It can be found, from Eqs. (\ref{filter1})-(\ref{form}), that the
coefficient $\det f_A\det f_B$ is equal to the minimum eigenvalue $\lambda
_{\min }^F$ of the associated operator $F_{AB}$. Thus we have 
\begin{equation}
P_f=\lambda _{\min }^FtrR(\rho ).  \label{probs}
\end{equation}

{\it Illustration}.---We now apply the results derived above to treat the
following four-parameters family of mixed states, 
\begin{equation}
\rho =p_1|\psi _1\rangle \langle \psi _1|+p_2|\psi _2\rangle \langle \psi
_2|+p_3|00\rangle \langle 00|+p_4|11\rangle \langle 11|.  \label{density}
\end{equation}
Here 
\begin{equation}
|\psi _1\rangle =\alpha |01\rangle -\beta |10\rangle ,\ |\psi _2\rangle
=\beta |01\rangle +\alpha |10\rangle  \label{state}
\end{equation}
and $\alpha $, $\beta $ are two positive real numbers satisfying $\alpha
^2+\beta ^2=1$. The four states $|\psi _1\rangle $, $|\psi _2\rangle $, $%
|00\rangle $ and $|11\rangle $ compose the orthogonal decomposition of $\rho 
$ and the corresponding probabilities are assumed to satisfy $p_1\geq p_2$
and $p_3\geq p_4$. Exploration of the states (\ref{density}) should be of
universal value: it comprises lower rank states as some of the coefficients $%
p_i$ reduce to zero, so the results acquired should predict the various
consequences mentioned above for these reduced cases. We demonstrate these
expectations in detail below.

A direct calculation gives the Wootters representation of the state $\rho $
with 
\begin{eqnarray}
|x_1\rangle &=&i\cos \theta \sqrt{p_1}|\psi _1\rangle -i\sin \theta \sqrt{
p_2}|\psi _2\rangle ,  \nonumber \\
|x_2\rangle &=&\sin \theta \sqrt{p_1}|\psi _1\rangle +\cos \theta \sqrt{ p_2}%
|\psi _2\rangle ,  \nonumber \\
|x_3\rangle &=&i\frac{\sqrt{2}}2(\sqrt{p_3}|00\rangle +\sqrt{p_4}|11\rangle ,
\nonumber \\
|x_4\rangle &=&\frac{\sqrt{2}}2(\sqrt{p_3}|00\rangle -\sqrt{p_4}|11\rangle ,
\label{wootx}
\end{eqnarray}
and their tilde inner products are 
\begin{eqnarray}
\lambda _1 &=&[\alpha ^2\beta ^2(p_1-p_2)^2+p_1p_2]^{1/2}+\alpha \beta
(p_1-p_2),  \nonumber \\
\lambda _2 &=&[\alpha ^2\beta ^2(p_1-p_2)^2+p_1p_2]^{1/2}-\alpha \beta
(p_1-p_2),  \nonumber \\
\lambda _3 &=&\lambda _4=\sqrt{p_3p_4}.  \label{tipro}
\end{eqnarray}
The parameter $\theta $ in (\ref{wootx}) is given by 
\begin{equation}
\theta =\arctan \frac{\sqrt{p_1p_2}(\alpha ^2-\beta ^2)}{[\alpha ^2\beta
^2(p_1-p_2)^2+p_1p_2]^{1/2}+\alpha \beta (p_1+p_2)}.  \label{ceta}
\end{equation}
Provided the presumption that the considered state $\rho $ has nonzero
entanglement of formation (thus nonzero concurrence), the parameters $\alpha 
$, $\beta $, and $p_i$s should satisfy 
\begin{equation}
C(\rho )=2[\alpha \beta (p_1-p_2)-\sqrt{p_3p_4}]>0.  \label{concur}
\end{equation}
Now according to the results derived in the preceding section, the local
operation of the optimal transformation for the rank-$4$ state $\rho $ can
be uniquely determined. The factorized form of the associated operator $%
F_{AB}$ can be obtained by a step-by-step calculation from the proposed
approach. The final form of it can be shown as 
\begin{equation}
F_{AB}=k^2(\frac 1{k^2}\sqrt{\frac{p_4}{p_3}}|0\rangle \langle 0|+|1\rangle
\langle 1|)\otimes (|0\rangle \langle 0|+\frac 1{k^2}\sqrt{\frac{p_3}{p_4}}%
|1\rangle \langle 1|),  \label{result}
\end{equation}
where 
\begin{equation}
k\equiv \sqrt{\frac 2{\lambda _1}}(\sqrt{p_1}\alpha \cos \theta -\sqrt{p_2}%
\beta \sin \theta ).  \label{k}
\end{equation}
The local filters $f_A$ and $f_B$ can be easily determined from Eqs. (\ref
{filter1})-(\ref{form}). For example, as the parameters satisfying $\sqrt{%
p_4/p_3}\leq \sqrt{p_3/p_4}\leq k^2$, they can be shown as 
\begin{equation}
f_A=\frac 1k(\frac{p_4}{p_3})^{1/4}|0\rangle \langle 0|+|1\rangle \langle
1|,f_B=|0\rangle \langle 0|+\frac 1k(\frac{p_3}{p_4})^{1/4}|1\rangle
\langle 1|,
\label{result1}
\end{equation}
and for the case $k^2\leq \sqrt{p_4/p_3}\leq \sqrt{p_3/p_4}$, they are 
\begin{equation}
f_A=|0\rangle \langle 0|+k(\frac{p_3}{p_4})^{1/4}|1\rangle \langle 1|,
f_B=k(\frac{p_4}{p_3})^{1/4}|0\rangle \langle 0|+|1\rangle \langle 1|.
\label{result2}
\end{equation}
The probability of the transformation is given by 
\begin{equation}
P_f=\frac 2{k^2}\left\{ [\alpha ^2\beta ^2(p_1-p_2)^2+p_1p_2]^{1/2}+\sqrt{%
p_3p_4}\right\}  \label{prob1}
\end{equation}
or 
\begin{equation}
P_f=2k^2\left\{ [\alpha ^2\beta ^2(p_1-p_2)^2+p_1p_2]^{1/2}+\sqrt{p_3p_4}%
\right\}  \label{prob2}
\end{equation}
corresponding to the above two cases, respectively. The transformed Bell
diagonal state can be worked out from Eq. (\ref{transtate}). As $p_2=0$, it
can be shown simply 
\begin{equation}
\rho ^{\prime }=\frac 1N[2p_1\alpha \beta |\Psi _{-}\rangle \langle \Psi
_{-}|+\sqrt{p_3p_4}(|\Phi _{+}\rangle \langle \Phi _{+}|+|\Phi _{-}\rangle
\langle \Phi _{-}|)],  \label{finstate}
\end{equation}
where the normalized factor $N=2p_1\alpha \beta +2\sqrt{p_3p_4}$ and $%
|\Psi _{-}\rangle $ and $|\Phi _{\pm }\rangle $ are Bell states given by 
\begin{eqnarray}
|\Psi _{-}\rangle &=&\frac{\sqrt{2}}2(|01\rangle -|10\rangle ),  \nonumber \\
|\Phi _{\pm }\rangle &=&\frac{\sqrt{2}}2(|00\rangle \pm |11\rangle ).
\label{bellst}
\end{eqnarray}
The state $\rho ^{\prime }$ has the maximal extractable concurrence 
\begin{equation}
C(\rho ^{\prime })=\frac{p_1\alpha \beta -\sqrt{p_3p_4}}{p_1\alpha \beta +%
\sqrt{p_3p_4}}=\left[ \frac{C(\rho )}{trR(\rho )}\right] _{p_2=0},
\label{concur1}
\end{equation}
which verifies the formula (12) presented in Ref. \cite{cen}.

Now let us give a discussion for the degenerate cases.

1. First consider the situation as the parameter $p_4$ approaches to zero
asymptotically and the limit of it corresponds to a rank $3$ density matrix
with $\lambda _3=0$ (or a rank $2$ state with $\lambda _2=0$ as $p_2=0$). In
such a case, the relation $\sqrt{p_4/p_3}\leq k^2\leq \sqrt{p_3/p_4}$ exists
and the local filters $f_A$ and $f_B$ can be obtained from Eq. (\ref
{result}) as
\begin{equation}
f_A=\frac 1k(\frac{p_4}{p_3})^{1/4}|0\rangle \langle 0|+|1\rangle \langle
1|, f_B=k(\frac{p_4}{p_3})^{1/4}|0\rangle \langle 0|+|1\rangle \langle 1|.
\label{case1f}
\end{equation}
One can see in the limit of $p_4\rightarrow 0$, $f_A$ and $f_B$ reduce to
projection operators of pure states of the two qubits, respectively, and the
probability of the manipulation
\begin{equation}
P_f=2\sqrt{\frac{p_4}{p_3}}\left\{ [\alpha ^2\beta
^2(p_1-p_2)^2+p_1p_2]^{1/2}+\sqrt{p_3p_4}\right\}   \label{case1p}
\end{equation}
becomes zero. This confirms the assertion that any two-qubit density matrix
with $\lambda _n=0$ cannot be transformed into Bell diagonal state
efficiently by invertible local actions.

2. Another notable situation is the rank $2$ case with $p_3=$ $p_4=0$. In
this case the expression given by (\ref{result}) appears to be not valid any
more. The analysis proposed previously predicts that there exist various
solutions for the local operator, and they can be obtained by establishing
the bases \{$|x_3\rangle ,|x_4\rangle $\} as 
\begin{equation}
|x_3\rangle =i(\tau _1|00\rangle +\tau _2|11\rangle ),\ |x_4\rangle =\tau
_1|00\rangle -\tau _2|11\rangle ,  \label{selec}
\end{equation}
where $\tau _1$ and $\tau _2$ are complex parameters and the multiplication
of them satisfies $\tau _1\tau _2=|\tau _1\tau _2|>0$. Similarly, the
factorizable form of the associated operator $F_{AB}$ can be worked out 
\begin{equation}
F_{AB}=k^2(\frac 1{k^2}\frac{|\tau _2|}{|\tau _1|}|0\rangle \langle
0|+|1\rangle \langle 1|)\otimes (|0\rangle \langle 0|+\frac 1{k^2}\frac{
|\tau _1|}{|\tau _2|}|1\rangle \langle 1|).  \label{rank2}
\end{equation}
In comparison with Eq. (\ref{result}), the difference is that the ratio $
|\tau _2|/|\tau _1|$ in (\ref{rank2}) can be an arbitrary positive number,
which implies the multiplicity of the local action. As the parameters are
chosen that $|\tau _1|=|\tau _2|$, the local action has the optimal
probability 
\begin{equation}
P_f=2[\alpha ^2\beta ^2(p_1-p_2)^2+p_1p_2]^{1/2}\times \min \{k^2,\frac 1{k^2%
}\}.  \label{prob}
\end{equation}

In conclusion, we have explored the optimal entanglement manipulation for a
single copy of two-qubit states. By virtue of a particular representation of
the two-qubit state, an explicit form of the LOCC action has been derived to
transform the state into Bell diagonal state and the probability of the
transformation was also given. The results display that the local operation
of such a process is unique for the density matrices with the rank $n=3$ and 
$4$, whereas it can be constructed variously for that with $n=2$.

\acknowledgements

This work was supported in part by the Postdoctoral Science Foundation and
the special funds for Major State Basic Research Project No. G001CB3095 of
China.


\begin{references}
\bibitem{note}  For the states with rank $n=4$, there exists $\langle x_4|
\tilde{x}_4\rangle \neq 0$. Since if $\langle x_4|\tilde{x}_4\rangle =0$,
the state $|\tilde{x}_4\rangle $ shall be orthogonal to all the four
linearly-independent states $|x_i\rangle $ ($i=1,\cdots ,4$). This obviously
cannot happen in the $4$-dimensional Hilbert space.
\end{references}
\end{document}